# Gravitational Wave (GW) Classification, Space GW Detection Sensitivities and AMIGO (Astrodynamical Middle-frequency Interferometric GW Observatory)

*Wei-Tou* Ni[1]

Center for Gravitation and Cosmology (CGC), Department of Physics, National Tsing Hua University, No. 101, Kuang Fu II Rd., Hsinchu, Taiwan, 300 ROC

**Abstract.** After first reviewing the gravitational wave (GW) spectral classification. we discuss the sensitivities of GW detection in space aimed at low frequency band (100 nHz–100 mHz) and middle frequency band (100 mHz–10 Hz). The science goals are to detect GWs from (i) Supermassive Black Holes; (ii) Extreme-Mass-Ratio Black Hole Inspirals; (iii) Intermediate-Mass Black Holes; (iv) Galactic Compact Binaries; (v) Stellar-Size Black Hole Binaries; and (vi) Relic GW Background. The detector proposals have arm length ranging from 100 km to $1.35\times10^9$ km (9 AU) including (a) Solar orbiting detectors and (b) Earth orbiting detectors. We discuss especially the sensitivities in the frequency band 0.1-10 μHz and the middle frequency band (0.1 Hz–10 Hz). We propose and discuss AMIGO as an Astrodynamical Middle-frequency Interferometric GW Observatory.

## 1 Classification and Spectral Sensitivities

With LIGO's direct detection of the binary black hole merger events [1, 2, 3], we have been fully ushered into the age of Gravitational Wave (GW) astronomy. Detection efforts over all GW frequency bands from cosmological frequency band (1 aHz–10 fHz) to ultra-high frequency band (over 1 THz) have been vigorously exerted (See, e.g. [4]). In Table 1, we list the band ranges and the detection methods [4-6]. We have also plotted the GW detector sensitivities and GW source strengths on single diagrams with ordinates showing characteristic strain, strain power spectral density (psd) amplitude and normalized GW spectral energy density respectively in 2015 [4]. Currently we are updating these diagrams [7]. Fig. 1 shows the strain psd amplitude vs. frequency for various detectors and sources adapting from the corresponding figures of Refs. [4] and [7]. For detailed explanation of the plot, see [4, 7].

Presently, most GW detection efforts are spent in 4 bands – the high frequency band (10 Hz–100 kHz), the upper part (10 μHz–0.1 Hz) of the low frequency band, the very low frequency band (nano-Hz band, 300 pHz –100 nHz) and the extremely low (Hubble) frequency band (cosmological band or CMB band; 1 aHz–10 fHz). As can be seen in Fig. 1, there are 3 regions which are poor in the near-future projected sensitivities adjacent to these 4 bands: (i) the middle frequency band, (ii) the lower part (100 nHz–10 μHz) of the low frequency band and (iii) the ultralow frequency band (10 fHz–300 pHz). To possibly increase the sensitivity in the frequency band 0.1-10 μHz, Super-ASTROD with arm length of 9 AU has been proposed [8]. To have significant sensitivity in the frequency band 0.1–10 Hz and yet to be a first-generation candidate for space GW missions, we propose a middle-frequency GW mission AMIGO (Astrodynamical Middle-frequency Interferometric GW Observatory) with arm length 10,000 km and discuss the concept in section 3 after a review on space GW detection sensitivities in section 2.

---
[1] weitou@gmail.com

**Table 1.** Frequency classification of gravitational waves and their detection method [4-6]

| Frequency band | Detection method |
| --- | --- |
| Ultra high frequency band: above 1 THz | Terahertz resonators, optical resonators, and magnetic conversion detectors |
| Very high frequency band: 100 kHz–1 THz | Microwave resonator/wave guide detectors, laser interferometers and Gaussian beam detectors |
| High frequency band (audio band)*: 10 Hz–100 kHz | Low-temperature resonators and ground-based laser-interferometric detectors |
| Middle frequency band: 0.1 Hz–10 Hz | Space laser-interferometric detectors of arm length 100 km − 60,000 km, atom and molecule interferometry, optical clock detectors |
| Low frequency band (milli-Hz band)[†]: 100 nHz–0.1 Hz | Radio Doppler tracking of spacecraft, space laser-interferometric detectors of arm length longer than 60,000 km, optical clock detectors |
| Very low frequency band (nano-Hz band): 300 pHz – 100 nHz | Pulsar timing arrays (PTAs) |
| Ultralow frequency band: 10 fHz–300 pHz | Astrometry of quasars and their proper motions |
| Extremely low (Hubble) frequency band (cosmological band): 1 aHz–10 fHz | Cosmic microwave background experiments |
| Beyond Hubble-frequency band: below 1 aHz | Through the verifications of inflationary/primordial cosmological models |

*The range of audio band (also called LIGO band) normally goes only to 10 kHz.
[†]The range of milli-Hz band is 0.1 mHz to 100 mHz.

## 2 SPACE GW detection sensitivities

GW detection in space aimed at low frequency band (100 nHz – 100 mHz) and middle frequency band (100 mHz – 10 Hz). Its scientific goals are to detect GWs from (i) Supermassive Black Hole Coalescences; (ii) Extreme-Mass-Ratio Black Hole Inspirals; (iii) Intermediate-Mass Black Hole Coalescences; (iv) Galactic Compact Binaries, (v) Stellar-size Black Hole Binary Inspirals, and (vi) Relic GW Background.

The main technological requirements of GW detection in space are (i) drag-free requirement; and (ii) requirement of measuring relative distance variation or relative velocity variation. LISA Pathfinder (LPF) launched on 3 December 2015, has achieved not only the drag-free requirement goal of this technology demonstration mission, but also has completely met the more stringent LISA drag-free demand [9-11]. In short, LISA Pathfinder has successfully demonstrated the first generation drag-free technology requirement for space detection of GWs.

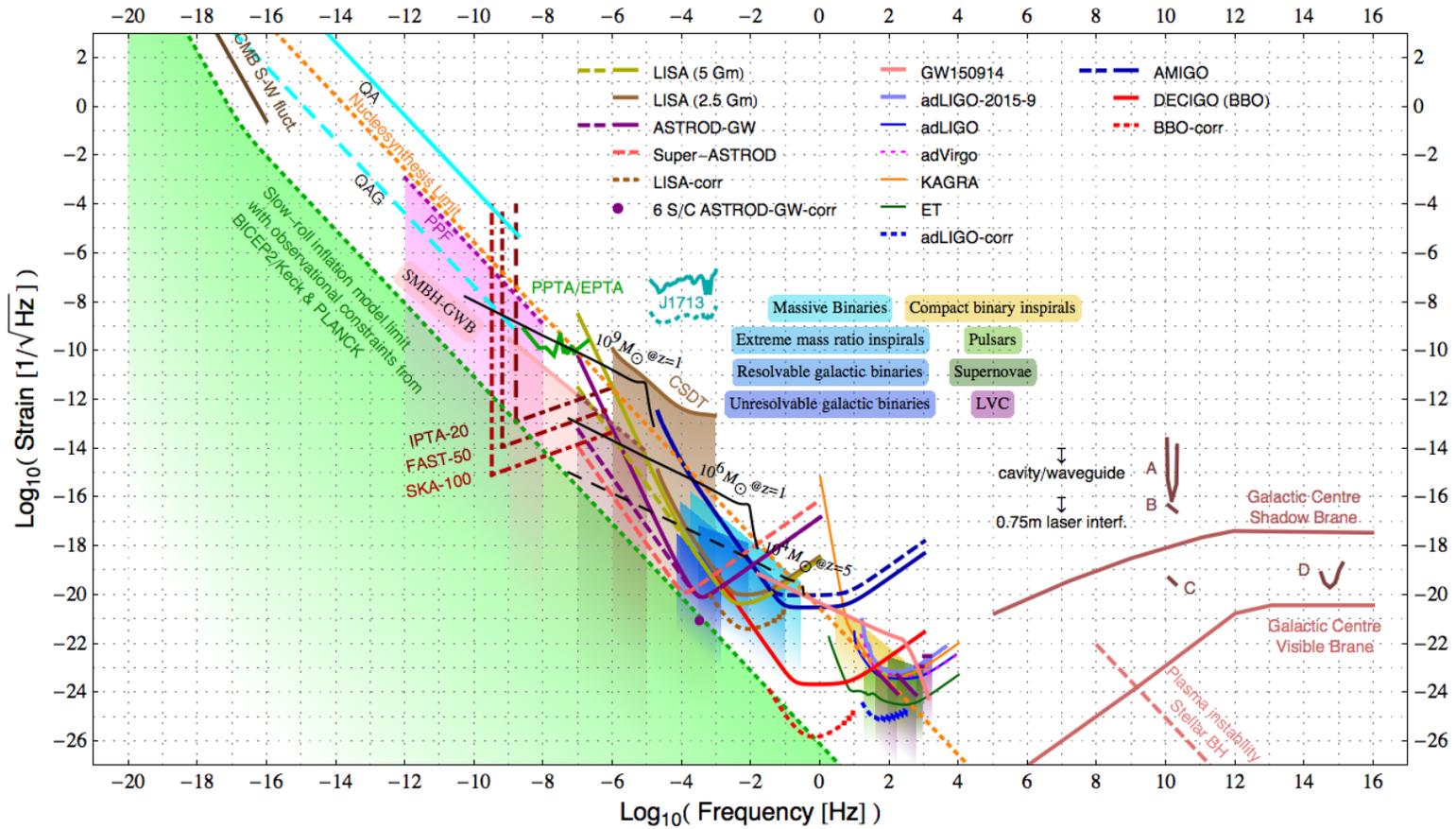

**Fig. 1.** Strain psd amplitude vs. frequency for various GW detectors and sources. [QA: Quasar Astrometry; QAG: Quasar Astrometry Goal; PPF: Pulsar Parameter Fitting; LVC: LIGO-Virgo Constraints; CSDT: Cassini Spacecraft Doppler Tracking; SMBH-GWB: Supermassive Black Hole-GW Background.]

The requirement of measuring relative distance variation or relative velocity variation is in terms of spectral strain sensitivity. For space GW detection, the first-generation requirement is around $10^{-20}$ Hz$^{-1/2}$ sensitivity for measurement of strain psd amplitude. For measurement using unequal-arm laser interferometry, the requirement on laser stabilization is similar. However, the present laser stabilization has not reached this kind of stability. One needs to match the two optical paths using Time Delay Interferometry (TDI) to lessen the stability requirement. For TDI configurations and their numerical simulations for various missions, see Tinto and Dhurandhar [12], Wang and Ni [13] and references therein. Experimental demonstration of TDI in laboratory for LISA worked out in 2010-2012 (Vine et al. [14], Mirtyk et al. [15]).

In space, Michelson type interferometry invariably involve large distances. The laser power received at the far end of the optical link is weak. To continue the optical path as required by TDIs, one needs to amplify it. The way of amplification is to track the optical phase of the incoming weak light with the local laser oscillator by optical phase-locking. At National Tsing Hua University, 2 pW weak-light homodyne phase-locking with 0.2 mW local oscillator has been demonstrated (Liao et al. [16, 17]). In JPL (Jet Propulsion Laboratory), Dick et al. [18] have achieved offset phase locking of local oscillator to 40 fW incoming laser light. More recently, Gerberding et al. [19] and Francis et al. [20] have phase-locked and tracked a 3.5 pW weak light signal and a 30 fW weak light signal respectively at reduced cycle slipping rate. For LISA, 85 pW weak-light phase locking is required. For ASTROD-GW, 100 fW weak-light phase locking is required. Hence, the weak level of these weak-light power requirements has achieved. In the future, the frequency-tracking, modulation-demodulation and coding-decoding needs development to make it a mature technology. This is also important for deep space CW (Continuous Wave) optical communication.

As shown in Fig. 1, typical frequency sensitivity spectrum of strain psd amplitude for space GW detection consists of three regions, the acceleration/local gravity gradient/vibration noise dominated region, the shot noise (flat for current space detector projects like LISA in strain psd) dominated region, if any, and the antenna response restricted region. The detector sensitivity in the lower frequency region is constrained by vibration, acceleration noise or gravity-gradient noise. The detector sensitivity of the higher frequency part is constrained by antenna response (or storage time). In a power-limited design, sometimes there is a middle flat region in which the sensitivity is limited by the photon shot noise. [21-24]

The shot noise sensitivity in the strain for GW detection is inversely proportional to $P^{1/2}L$ with $P$ the received power and $L$ the distance or arm length. Since $P$ is inversely proportional to $L^2$ and $P^{1/2}L$ is constant, this sensitivity limit is independent of the distance. For 1-2 W emitting power, the limit is around $10^{-20}$–$10^{-21}$ Hz$^{-1/2}$ (depending on telescope diameter/laser beam divergence). As noted in the LISA study [21], making the arms longer shifts the time-integrated sensitivity curve to lower frequencies while leaving the bottom of the curve at the same level. Hence, ASTROD-GW with longer arm length has better sensitivity at lower frequency. e-LISA, ALIA, TAIJI, and GW interferometers in Earth orbit have shorter arms and therefore have better sensitivities at higher frequency.

In Fig. 1, we plot the sensitivity curves for LISA (5 Gm), LISA (2.5 Gm, i.e. the new LISA), ASTROD-GW, BBO, DECIGO, AMIGO and Super-ASTROD. The sensitivity curves for LISA (5 Gm), ASTROD-GW, BBO and DECIGO are taken from Fig. 3 of [4] and references therein. Others are from [7] and references therein. Fig. 2 is a blowup of Fig. 1 restricted to the frequency range $10^{-10}$–$10^4$ Hz to show the space frequency regions and the 2 neighboring regions. Section 2.1 discusses new LISA (2.5 Gm) and gives its sensitivity equation. Section 2.2 does it for Super-ASTROD.

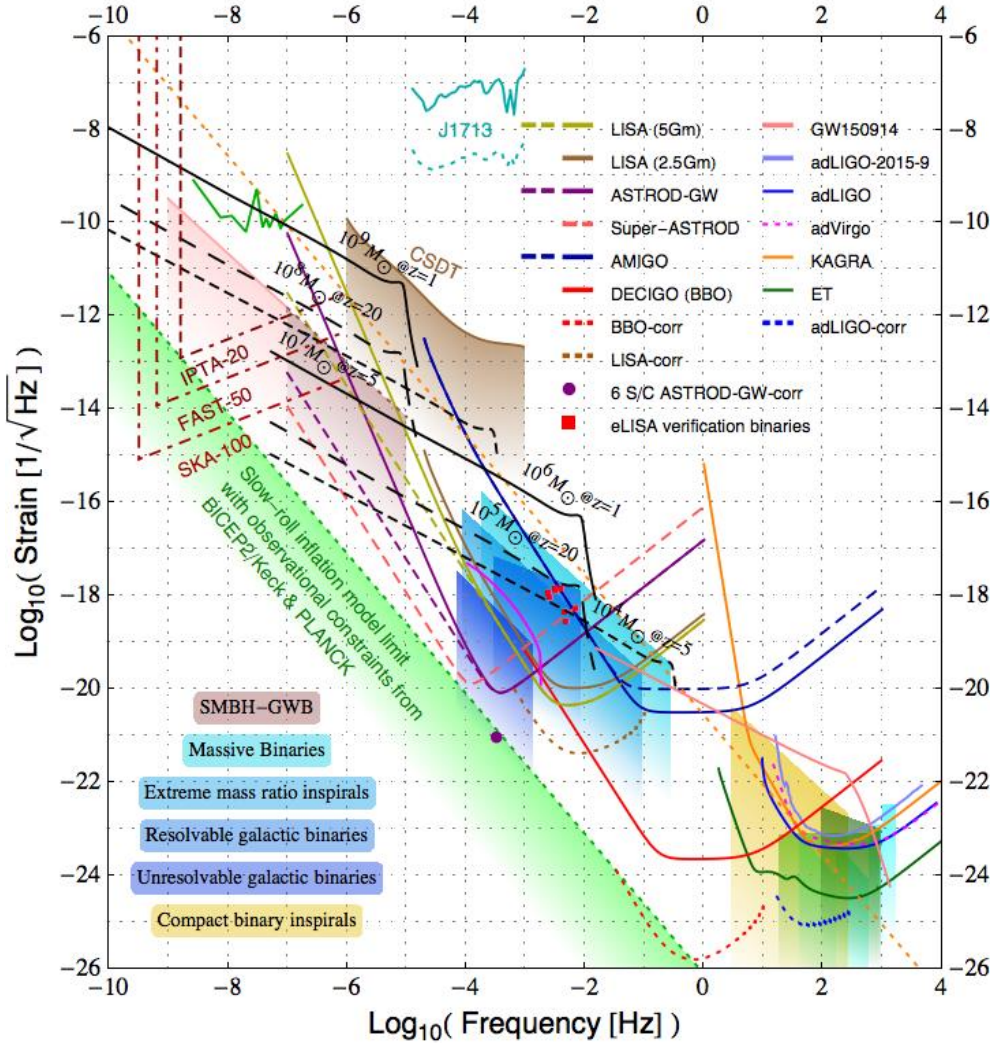

**Fig. 2.** Strain power spectral density (psd) amplitude vs. frequency for various GW detectors and GW sources. The black lines show the inspiral, coalescence and oscillation phases of GW emission from various equal-mass black-hole binary mergers in circular orbits at various redshift: solid line, $z = 1$; dashed line, $z = 5$; long-dashed line $z = 20$. See text of [24] for more explanation. The strain psd amplitude of GW150914 is calculated from its characteristic amplitude in Figure 1 of [25] using standard formula. The AMIGO design sensitivity is in solid blue while AMIGO baseline sensitivity is in dashed blue. The two curves merge together at lower frequency in the figure. [CSDT: Cassini Spacecraft Doppler Tracking; SMBH-GWB: Supermassive Black Hole-GW Background.]

## 2.1 LISA (2.5 Gm)

A new LISA proposal (Amaro-Seoane et al. [11]) was submitted to ESA on January 13th in response to the call for missions for the L3 slot in the Cosmic Vision Programme. On 20 June 2017, ESA announced the news that "The LISA trio of satellites to detect gravitational waves from space has been selected as the third large-class mission in ESA's Science programme (ESA 2017)." The basic concept is the same as the original LISA, but with arm length down-scaled to 2.5 Gm from 5 Gm. To distinguish this selected mission proposal from the original one or the NGO/eLISA, we call it LISA (2.5 Gm or new LISA in case of possible ambiguity. Quoting from the proposal [11]:

"The observatory will be based on three arms with six active laser links, between three identical spacecraft in a triangular formation separated by 2.5 million km. Continuously operating heterodyne laser interferometers measure with pm Hz$^{-1/2}$ sensitivity in both directions along each arm, using well-stabilized lasers at 1064 nm delivering 2 W of power to the optical system. Using technology proven in LISA Pathfinder, the Interferometry Measurement System is using optical benches in each spacecraft constructed from an ultra-low expansion glass-ceramic to minimize optical path length changes due to temperature fluctuations. 30 cm telescopes transmit and receive the laser light to and from the other spacecraft. Three independent interferometric combinations of the light travel time between the test masses are possible, allowing, in data processing on the ground, the synthesis of two virtual Michelson interferometers plus a third null-stream, or "Sagnac" configuration." These two virtual Michelson interferometers are two TDIs. They could be two out of three TDI configurations X, Y and Z if they satisfy the noise requirement.

The new LISA design sensitivity is in [10, 11]. A simple analytical approximation of the design sensitivity is in Petiteau et al. [10] and used by Cornish and Robson [26]:

$$S_{Ln}^{1/2}(f) = (20/3)^{1/2} (1/L_L) \times [(1 + (f/(1.29 f_L))^2)]^{1/2} \times [(S_{Lp} + 4 S_a/(2\pi f)^4)]^{1/2} \text{ Hz}^{-1/2}, \quad (1)$$

over the frequency range 20 μHz $< f <$ 1 Hz. Here $L_L$ = 2.5 Gm is the LISA arm length, $f_L = c/(2\pi L_L)$ is the LISA arm transfer frequency, $S_{Lp} = 8.9 \times 10^{-23}$ m$^2$ Hz$^{-1}$ is the white position noise, and

$$S_a(f) = 9 \times 10^{-30} \ [1 + (10^{-4} \text{ Hz}/f)^2 + 16 \ (2 \times 10^{-5} \text{ Hz}/f)^{10}] \text{ m}^2 \text{ s}^{-4} \text{ Hz}^{-1}, \quad (2)$$

is the colored acceleration noise level. This new LISA design sensitivity curve shows in both Fig. 1 and Fig. 2.

## 2.2 Super-ASTROD

Super-ASTROD (Super Astrodynamical Space Test of Relativity using Optical Devices) is a mission concept with four spacecraft in 5.2 AU orbits together with an Earth–Sun L1/L2 spacecraft ranging optically with one another to probe GWs with frequencies 0.1 μHz–1 mHz, to test fundamental laws of spacetime and to map the outer-solar-system mass distribution and dynamics [8]. In this paper, we address mainly the GW sensitivity of Super-ASTROD for enhancing the GW detection in the sensitivity gap 100 nHz−10 μHz. With three spacecraft in Jupiter-like solar orbits of radius 5.2 AU and period of about 12 year, the desired baseline mission lifetime would be 25 years or more. The orbits of these 3 spacecraft are near the Sun-Jupiter L3, L4, L5 points respectively to form an ASTROD-GW-like configuration [23, 24]. The Super-ASTROD spacecraft configuration will be inclined to the Jupiter orbit plane by 1-3°. The angular precession of the spacecraft configuration will be twice the Jupiter orbiting angular velocity around the Sun, i.e. about 6 years in period. In 6 years, the angular position of quasi-monochromatic GW sources will be resolved.

Due to its large extension, Super-ASTROD is a second-generation GW mission concept. To set its sensitivity goal, we review the GW sensitivity of ASTROD-GW [23, 24] in Fig. 1.

For ASTROD-GW, our goal on the instrumental strain psd amplitude is

$$S_{An}^{1/2}(f) = (1/L_A) \times \{[(1 + 0.5 \ (f/f_A)^2)] \times S_{Ap} + [4 S_a/(2\pi f)^4]\}^{1/2} \text{ Hz}^{-1/2}, \quad (3)$$

over the frequency range of 100 nHz < $f$ < 1 Hz. Here $L_A = 260 \times 10^9$ m is the ASTROD-GW arm length, $f_A = c / (2\pi L_A)$ is the ASTROD-GW arm transfer frequency, $S_a = 9 \times 10^{-30}$ m² s⁻⁴ Hz⁻¹ is the white acceleration noise level (the same as that for original LISA [21]), and $S_{Ap} = 10816 \times 10^{-22}$ m² Hz⁻¹ is the (white) position noise level due to laser shot noise which is 2704 (=52²) times that for original LISA [23]. The corresponding noise curve for the ASTROD-GW instrumental noise psd amplitude $^{(MLDC)}S_{An}^{1/2}(f)$ with the same "reddening" factor as specified in MLDC program is

$$^{(MLDC)}S_{An}^{1/2}(f) = (1/L_A) \times \{[(1 + 0.5 \ (f/f_A)^2)] \times S_{Ap} + [1 + (10^{-4}/f)^2] \ (4S_a/(2\pi f)^4)\}^{1/2} \ \text{Hz}^{-1/2}, \quad (4)$$

over the frequency range of 100 nHz < $f$ < 1 Hz. The one without reddening factor are shown with dashed line in the lower frequency part.

For Super-ASTROD, our goal on the instrumental strain noise psd amplitude is

$$S_{Sn}^{1/2}(f) = (1/L_S) \times \{[(1 + 0.5 \ (f/f_S)^2)] \times S_{Sp} + [4S_a/(2\pi f)^4]\}^{1/2} \ \text{Hz}^{-1/2}, \quad (5)$$

over the frequency range of 100 nHz < $f$ < 1 Hz. Here $L_S = 1350 \times 10^9$ m (9 AU) is the nominal Super-ASTROD arm length, $f_S = c / (2\pi L_S)$ is the Super-ASTROD arm transfer frequency, $S_a = 9 \times 10^{-30}$ m² s⁻⁴ Hz⁻¹ is the white acceleration noise level (the same as that for original LISA and ASTROD-GW), and $S_{Sp} = 291600 \times 10^{-22}$ m² Hz⁻¹ is the (white) position noise level due to laser shot noise which is 72900 (=270²) times that for original LISA.

## 3 AMIGO

A discussion of ground-based GW detector concepts to extend the present ground-based interferometers detection spectral range, i.e., the high-frequency GW band 10 Hz–100 Hz to middle-frequency band 0.1–10 Hz together with the plethora of potential astrophysical sources in this band is given in Harms et al. [27]. Harms et al. examine the potential sensitivity of three detection concepts (atom interferometers, torsion bar antennas and Michelson interferometers), estimate for their event rates and thereby, the sensitivity requirements for these detectors. They find that the scientific payoff from measuring astrophysical gravitational waves in this frequency band is great. However, although they find no fundamental limits to the detector sensitivity in this band, the remaining technical limits will be extremely challenging to overcome. In this conference, Ho Jung presents a middle-frequency GW detector concept SOGRO (Superconducting Omni-directional Gravitational Radiation Observatory) on Earth [28, 29]. The Newtonian-noise cancellation from infrasound and seismic surface fields is very challenging [30].

In this paper, we propose a *first-generation* middle-frequency mission concept AMIGO: Astrodynamical Middle-frequency Interferometric GW Observatory with the following specification:

Arm length: 10,000 km (or a few times of this)

Laser power: 2 – 10 W

Acceleration noise: same as LPF has already achieved

Orbits and formation: 4 options (all LISA-like formations):

    (i) Earth-like solar orbits (3-20 degrees behind the Earth orbit)

    (ii) 600,000 km high orbit formation around the Earth

    (iii) 100,000 km-250,000 high orbit formation around the Earth

    (iv) near Earth-Moon L4 (or L5) halo orbit formation

The Scientific Goals of AMIGO are: to bridge the spectra gap between first-generation high-frequency and low-frequency GW sensitivities; Detecting intermediate mass BH coalescence; Detecting inspiral phase and predict time of binary black hole coalescence together with neutron star coalescence for ground interferometers; Detecting compact binary inspirals for studying stellar evolution and galactic population.

For AMIGO, our baseline on the noise psd amplitude assuming 2 W laser power, 30 cm telescopes and same acceleration noise as new LISA is:

$$S_{\text{AMIGOn}}^{1/2}(f) = (20/3)^{1/2}(1/L_{\text{AMIGO}}) \times [(1+(f/(1.29 f_{\text{AMIGO}}))^2)]^{1/2} \times [(S_{\text{AMIGOp}} + 4 S_a/(2\pi f)^4)]^{1/2} \text{Hz}^{-1/2}, \quad (6)$$

over the frequency range of 20 μHz $< f <$ 1 kHz. Here $L_{\text{AMIGO}} = 0.01 \times 10^9$ m is the AMIGO arm length, $f_{\text{AMIGO}} = c/(2\pi L_{\text{AMIGO}})$ is the AMIGO arm transfer frequency, $S_{\text{AMIGOp}} = 1.424 \times 10^{-28}$ m$^2$ Hz$^{-1}$ is the (white) position noise level due to laser shot noise which is $16 \times 10^{-6}$ (=0.004$^2$) times that for new LISA. $S_a(f)$ is the same colored acceleration noise level in (2). The AMIGO baseline sensitivity (6) is plotted as AMIGO dashed curve in both Fig. 1 and Fig. 2.

Since power and lower shot noise is crucial in reach better sensitivity in middle part of the sensitivity curve, we use 10 W laser power and 36 cm ϕ as our design values of the AMIGO mission concept to gain a factor of 10 [≈ (10/2)×(36/30)$^4$] for shot noise design sensitivity. The AMIGO design sensitivity (6) is plotted as AMIGO solid curve in both Fig. 1 and Fig. 2 by using $S_{\text{AMIGOp}} = 0.1424 \times 10^{-28}$ m$^2$ Hz$^{-1}$.

In Fig. 1 and Fig. 2, the strain psd amplitude of GW150914 is calculated from its characteristic amplitude in Figure 1 of [25] using standard formula of conversion. AMIGO with either baseline sensitivity or design sensitivity would detect the inspiral phase of GW150914 and predict the coalescence time for the benefit of doing multi-messenger astronomy. However, the design sensitivity has better coverage in detecting the inspiral phase of neutron star coalescence events.

The numerical TDIs for AMIGO would be easier to design compared to new LISA due to AMIGO's shorter arm length. X, Y, Z TDI configurations are well suited for AMIGO. However, experimental requirement on TDI is more stringent and needs developments. More studies on the mission concept and various choices of orbit design are under preparation [31].

## 4 Discussion and outlook

LIGO's measurement of GWs from black hole coalescence fully ushered us into the era of GW astronomy. LISA space GW mission will explore the GW sources of large part of our Universe with large S/N ratio in the low-frequency part. On the even lower frequency side, PTAs are actively looking for GWs. Between the frequency bands sensitive to PTAs and to first generation space GW detector, Super-ASTROD is discussed as a second-generation space detector concept to have better sensitivity in the interim frequency band to explore GW sources.

Currently, a number of detection methods are proposed and under active research to bridge the middle-frequency band gap between Earth-based and space-borne GW observations with important science goals. In this band, technical limits will be extremely challenging to overcome for Earth-based due to Newtonian noises. In this paper, we propose a first-generation middle-frequency mission concept AMIGO with 10,000 km arm length. The technical readiness level is high. The sensitivity is good to reach science goals considered in the last section.

If a pathfinder mission is desired with 2-spacecraft demonstration of ranging in the solar-system for a LISA-like mission, the case with 2-5 degrees lagging behind the Earth of the first orbit choice in the last section could be considered. Just take one arm of this AMIGO case, it would be good to test many things in the solar system: deployment, both radio and laser communications, noise budget, and drag-free system together with a concentrated effort on distance metrology. It might be simpler than go to L1 or L2 Sun-Earth Lagrange point.

I would like to thank Ling-Wei Luo for his help in drawing Figure 1 and Figure 2.